\documentclass[%
prd%
 ,twocolumn%
 ,secnumarabic%
,amssymb, amsmath,nobibnotes, aps, prl]{revtex4}
\usepackage{graphicx}
\expandafter\ifx\csname package@font\endcsname\relax\else
 \expandafter\expandafter
 \expandafter\usepackage
 \expandafter\expandafter
 \expandafter{\csname package@font\endcsname}%
\fi

\begin{document}

\title{Parametric signal amplification to create a stiff optical bar}
\author{K. Somiya}
\email{somiya@phys.titech.ac.jp}
\author{J. Kato}
\author{K. Yano}
\affiliation{Department of Physics, Tokyo Institute of Technology}
\author{N. Saito}
\affiliation{Department of Physics, Ochanomizu University}
\date{February 2014}

\begin{abstract}
An optical cavity consisting of optically trapped mirrors makes a resonant bar that can be stiffer than diamond~{\cite{MIT}\cite{Matsumoto}}. A limitation of the stiffness arises in the length of the optical bar as a consequence of the finite light speed. High laser power and light mass mirrors are essential for realization of a long and stiff optical bar that can be useful for example in the gravitational-wave detector aiming at the observation of a signal from neutron-star collisions, supernovae, etc. In this letter, we introduce a parametric signal amplification scheme that realizes the long and stiff optical bar without the need to increase the laser power.
\end{abstract}

\maketitle

\section{Overview}

A Fabry-Perot cavity forms a standing wave optical resonator for light waves. With the resonating mode slightly detuned from the input carrier mode, a mirror motion introduces an amplitude modulation on the carrier light to drive the mirrors via radiation pressure and creates an optical bar~\cite{BC2}. Susceptibility of the cavity to the mirror motion is enhanced at the bar resonance ({\it optical spring}) and suppressed at frequencies below the resonance ({\it optical bar}). In an interferometric gravitational-wave detector~\cite{GW}, the optical spring is profitable to increase the signal-to-noise ratio for certain astronomical sources. In a cold-damping experiment~\cite{cold}, the optical spring is essential to realize a harmonic oscillator with low thermal fluctuations. For the both cases, it is important to increase the spring frequency so that the measurement can be performed at frequencies high enough to be free from environmental disturbances. With the circulating power and the mass of the mirrors being fixed, the stiffness of the bar is limited by the length of the cavity. This is due to the delay of the mirror position information to be delivered by the circulating light. While the cavity length is not important in the cold-damping experiment, the cavity has to be as long as possible for the gravitational-wave detector where the signal is proportional to the cavity length.

In this letter, we discuss the optical bar stiffness in terms of the cavity length and introduce a parametric signal amplification scheme to increase the spring frequency without increasing the circulating laser power. A remarkable point of this scheme is that we will be able not only to shift the spring frequency but also to create a spring at each frequency with frequency-dependent parametric amplification ({\it broadband optical spring}), which will be briefly explained in the end of the letter. We focus on the discussion for the gravitational-wave detector, but our scheme will be useful for the application in other optical systems.

\section{Optical spring}

\begin{figure}[t]
		\includegraphics[scale=0.5]{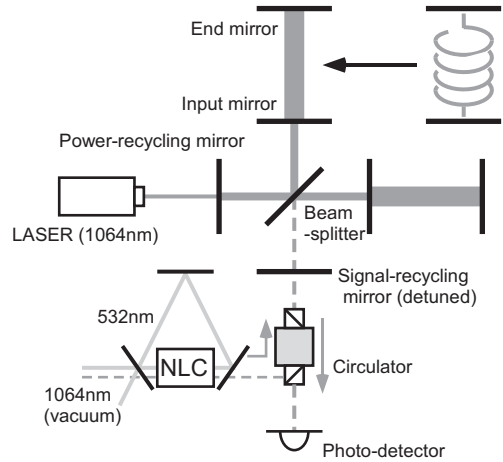}
		\caption{\label{fig:setup1}Configuration of a gravitational-wave detector with detuned signal recycling. The test masses in the arm cavities are connected by the optical bar. The vacuum fluctuation entering from the signal port is squeezed by the non-linear crystal (NLC) and a pump beam.}
\end{figure}

Figure~\ref{fig:setup1} shows a typical configuration of the interferometric gravitational-wave detector. It is based on the Michelson interferometer operated in the dark fringe, where all the input light except for some losses and gravitational-wave signals is reflected back toward the laser. The light split at the 50\,\% beamsplitter circulates in each arm cavity to increase the effective power and the signal. The light coming back from the beamsplitter toward the laser is recycled by another mirror located in between. Gravitational waves increase the distance of the two mirrors in one arm and decrease the distance in the other arm. The differential mode signal leaks through the interferometer toward the photo-detector. There is another mirror placed before the photo-detector and a cavity consisting of this mirror and the input mirrors of the arm cavities (signal-recycling cavity) is detuned from the resonance or anti-resonance of the carrier light to create an optical spring. This configuration is superior to a simple detuned Fabry-Perot cavity, which can also create an optical spring but at the same time decreases the circulating power. There is also a squeeze injection system depicted in Fig.~\ref{fig:setup1}, which will be explained later.

Gravitational waves induce phase modulation on the carrier light. If the signal-recycling cavity is tuned to the resonance or anti-resonance of the carrier, the phase signal reflected back to the interferometer is still in the phase modulation to the carrier light. With the detuning, the signal is in the mixed modulation and a fraction of the field in the amplitude modulation to the carrier light couples to the input carrier light to produce radiation pressure on the arm cavity mirrors. The mirror motion induces phase modulation and there is a loop of the signal production through the radiation pressure, which creates an optical spring.

\begin{figure}[htbp]
		\includegraphics[scale=0.4]{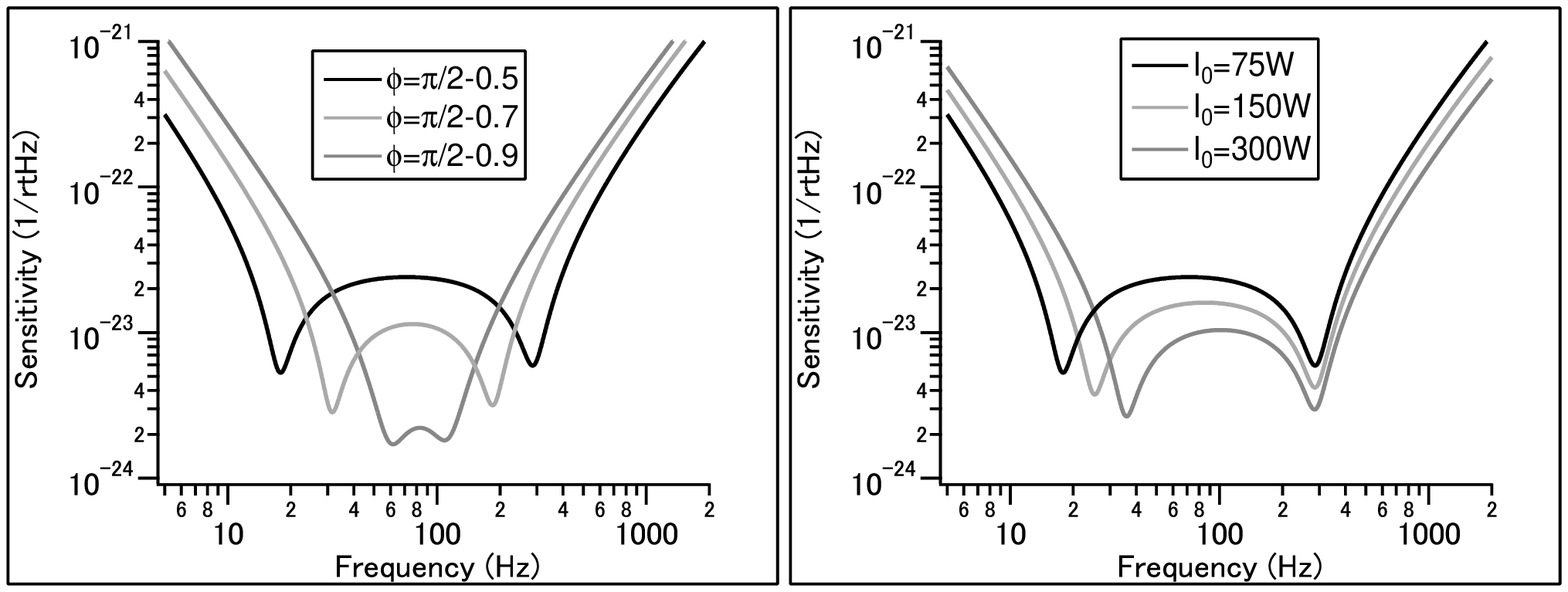}
		\caption{\label{fig:spectrum1}Quantum noise spectra of a detuned interferometer with different detune phases ({\it left}) and with different power levels ({\it right}).}
\end{figure}

The left panel of Fig.~\ref{fig:spectrum1} shows quantum noise spectra of the gravitational-wave detector with different detune phases. Hereafter, the input power, the power recycling gain, the arm cavity finesse, the signal-recycling mirror power reflectivity, and the detune phase are 75\,W, 11, 150, 85\,\%, and $\pi/2-0.5$\,rad, respectively, unless noted.
Origin of quantum noise is a vacuum fluctuation that enters the interferometer from the dark port. The noise level is given by the vacuum level divided by the signal strength. Each noise spectrum shows two dips: the lower dip is for the optical spring resonance and the upper dip is for the optical resonance of the coupled cavity with the detuned signal-recycling and the arm cavities. The two dips approach with the detune phase $\phi$ decreased from $\pi/2$. The highest end of the optical spring is the frequency where the two dips meet in the spectrum. The optical resonance of the coupled cavity is determined by the duration time of the signal field. The optical spring frequency cannot be higher than the optical resonance because it takes more time to deliver the information in the coupled cavity with such long duration time. 

The right panel of Fig.~\ref{fig:spectrum1} shows quantum noise spectra of the gravitational-wave detector with different circulating power. While the upper dip does not move, the lower dip moves upward with the increasing power, approximately by a factor proportional to the square root of the circulating power over the mirror mass~\cite{BC2}. We should note, however, that it is technically challenging to increase the circulating power and to lower the mass.

\section{Squeeze injection}

The increase of the circulating power is also a challenge in the gravitational-wave detector for the improvement of the sensitivity at high frequencies. One way to overcome the challenge is to use a squeeze injection technique. Instead of the coherent vacuum, a vacuum field that is squeezed in the phase quadrature is injected from the dark port of the interferometer so that the signal-to-noise ratio in a non-detuned interferometer can be improved as if the circulating power is increased. The squeezed vacuum can be generated with a non-linear crystal and a pump beam at the doubled frequency of the carrier light. The squeezing technique has been well developed and the squeezing factor is currently as high as 14\,dB~\cite{SQ}, which means that the quantum phase noise is improved as if the circulating power is made 25 times higher. The squeezing is, however, vulnerable to an optical loss in the interferometer. The optical loss does not only decrease the signal but also introduce a coherent vacuum field that deteriorates the squeezed vacuum.

\begin{figure}[htbp]
		\includegraphics[scale=0.4]{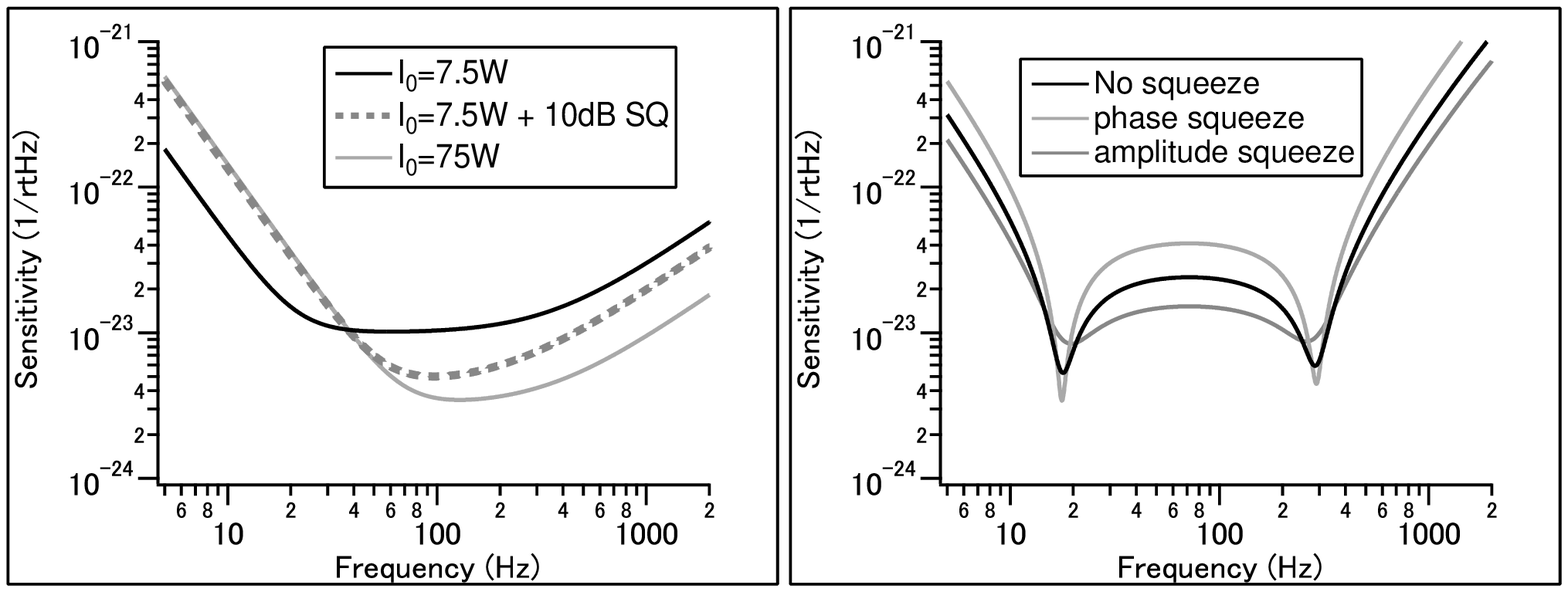}
		\caption{\label{fig:spectrum2}{\it Left}: Quantum noise spectra of a non-detuned interferometer with 10dB phase squeezing and without squeezing. The squeezing improves the sensitivity as if the input power is increased, besides the optical loss deteriorates the effect. {\it Right}: Quantum noise spectra of a detuned interferometer with 5dB phase/amplitude squeezing and without squeezing. The squeezing does not change the optical spring frequency but changes the steepness of the dips.}
\end{figure}

The squeezing is not equivalent to the increase of the power in a detuned interferometer. The sensitivity improvement in the non-detuned interferometer is due to the reduction of the vacuum field in the phase quadrature and thus does not affect the opto-mechanical dynamics. Figure~\ref{fig:spectrum2} shows the quantum noise spectra with and without the squeezing in a non-detuned interferometer (left panel) and in a detuned interferometer (right panel). The optical spring frequency does not change in the detuned interferometer. Instead, the steepness of the two dips increases with the squeezing in the detuned interferometer, which is due to the rotation of the squeeze quadrature around the dip frequencies. A nearly same situation could be realized by simply replacing the signal-recycling by a mirror with higher reflectivity. 

\section{Parametric amplification}

\begin{figure}[t]
		\includegraphics[scale=0.5]{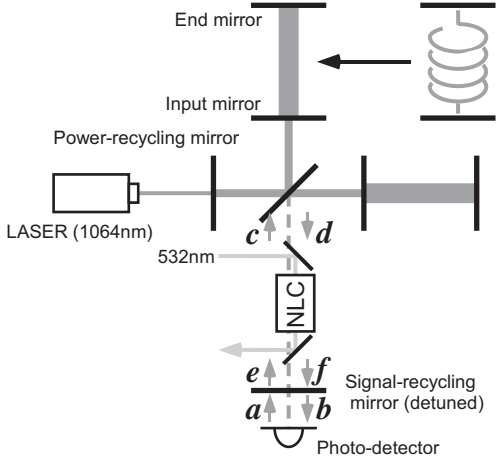}
		\caption{\label{fig:setup2}Configuration of a gravitational-wave detector with a detune signal recycling and a parametric amplifier. The vacuum fluctuation and the signal field coming out from the Michelson interferometer are amplified (anti-squeezed) by the non-linear crystal (NLC) and a pump beam.}
\end{figure}

Now we propose a new way to increase the optical spring frequency without increasing the duration time of the signal field or increasing the circulating power. Figure~\ref{fig:setup2} shows our setup. A squeezer is placed inside the signal-recycling cavity and is used as a signal amplifier. Unlike a regular squeezer, our squeezer increases the vacuum fluctuation in the phase quadrature as well as the gravitational-wave signal in the same quadrature. Without the signal-recycling mirror this would make no difference in the signal-to-noise ratio, but with the signal-recycling mirror that reflects back the amplified signal field the opto-mechanical dynamics of the interferometer changes as if the input laser power has increased. Since the signal amplification is instantaneous, the duration time problem is circumvented.

Mathematical description of the system is as follows. The electromagnetic field in the Heisenberg picture coming in or going out from the interferometer is given by
\begin{eqnarray}
E(t)=\sqrt{\frac{2\pi\hbar\omega_0}{{\cal A}c}}e^{-i\omega_0t}\!\!\int_0^\infty\!\!\!\!\left[a_+e^{-i\Omega t}+a_-e^{i\Omega t}\right]\frac{d\Omega}{2\pi}+\mathrm{H.C.},\nonumber
\end{eqnarray}
with $a_\pm$ the annihilation operator at $\omega_0\pm\Omega$. Here ${\cal A}$ is the effective cross sectional area of the laser beam and H.C. means Hermitian conjugate. As is performed in Ref.~\cite{BC2}, we introduce the two quadrature fields that express the amplitude quadrature (top element of a vector with subscript '1') that changes in-phase to the carrier light and the phase quadrature (bottom element with subscript '2') that changes with 90\,deg phase delay to the carrier light: $a_1=(a_++a_-)/\sqrt{2},\ a_2=(a_+-a_-)/\sqrt{2}i$.
These quadrature fields satisfy the commutation relations: $[a_1(\Omega),\,a_2^\dagger(\Omega')]=[a_2(\Omega),\,a_1^\dagger(\Omega')]=2\pi i\delta(\Omega-\Omega')$. Defining the electromagnetic fields as are shown in Fig.~\ref{fig:setup2}, we have
\begin{eqnarray}
&&\mbox{\boldmath $b$}=-r_s\mbox{\boldmath $a$}+t_s\mbox{\boldmath $f$}\ ,\ \ \ \ \ \ \ 
\mbox{\boldmath $e$}=r_s\mbox{\boldmath $f$}+t_s\mbox{\boldmath $a$}\ ,\nonumber\\
&&\mbox{\boldmath $f$}={\cal R}(\phi){\cal S}(s,\ \xi)\mbox{\boldmath $d$}\ ,\ \ \ 
\mbox{\boldmath $c$}={\cal R}(\phi)\mbox{\boldmath $e$}\ ,\label{eq:inputoutput}\\
&&\mbox{\boldmath $d$}=
\left (
\begin{array}{@{\,}cc@{\,}}
1&0\\
-{\cal K}&1
\end{array}\right )
\left (
\begin{array}{@{\,}c@{\,}}
c_1\\
c_2
\end{array}\right )e^{2i\beta}
+\alpha\left (
\begin{array}{@{\,}c@{\,}}
0\\
h
\end{array}\right )e^{i\beta}\ .\nonumber
\end{eqnarray}
Here $r_s$ and $t_s$ are amplitude reflectivity and transmittance of the signal recycling mirror, $S(s,\ \xi)$ is the squeezing matrix with a squeeze factor $s$ and squeeze angle $\xi$, $R(\phi)$ is the rotation matrix in the signal-recycling cavity with $\phi$ as the detune phase, and $h$ is the gravitational-wave signal in strain; also ${\cal K}$ is the opto-mechanical coupling coefficient of the Fabry-Perot Michelson interferometer, $\beta$ is the phase delay in the arm cavity, and $\alpha$ indicates the signal strength coming out from the Fabry-Perot Michelson interferometer:
\begin{eqnarray}
{\cal K}=\frac{8\omega_0I_0}{mL^2\Omega^2(\gamma^2+\Omega^2)},\ \ \tan{\beta}=\frac{\Omega}{\gamma},\ \ \alpha=\sqrt{\frac{2\omega_0I_0}{(\gamma^2+\Omega^2)\hbar}}\nonumber
\end{eqnarray}
where $\omega_0$ is the carrier light angular frequency, $I_0$ is the laser power at the beamsplitter, $L$ is the arm cavity length, $\Omega$ is the measurement angular frequency, $\gamma=Tc/4L$ is the cavity pole angular frequency with $T$ as the transmittance of the input mirror and c as the light speed, and $\hbar$ is the Planck's constant.

Solving the simultaneous equation~(\ref{eq:inputoutput}), one obtains the following input-output relation of the interferometer with the parametric amplifier:
\begin{eqnarray}
\mbox{\boldmath $b$}=
\frac{1}{2M}\left[\left (\!
\begin{array}{@{\,}cc@{\,}}
A_{11}&A_{12}\\
A_{21}&A_{22}
\end{array}\!\right )\!\!
\left (
\begin{array}{@{\,}c@{\,}}
a_1\\
a_2
\end{array}\!\right )e^{2i\beta}
+\left (\!
\begin{array}{@{\,}c@{\,}}
D_1\\
D_2
\end{array}\!\right )h\,e^{i\beta}\right]
\end{eqnarray}
where
\begin{eqnarray}
M&=&s-r_s\left[(1+s^2)\cos{2\phi}+{\cal K}\sin{2\phi}\right]e^{2i\beta}+sr_s^2e^{4i\beta},\nonumber\\
A_{11}&=&
(1+r_s^2)\left[(1+s^2)\cos{2\phi}+{\cal K}\sin{2\phi}\right]\nonumber\\
&&\ \ \ \ \ \ \ \ \ \ \ \ \ \ \ \ \ -4sr_s\cos{2\beta}-t_s^2(1-s^2),\nonumber\\
A_{12}&=&-2t_s^2\sin{\phi}\,[(1+s^2)\cos{\phi}+{\cal K}\sin{\phi}],\nonumber\\
A_{21}&=&2t_s^2\cos{\phi}\,[(1+s^2)\sin{\phi}-{\cal K}\cos{\phi}],\\
A_{22}&=&
(1+r_s^2)\left[(1+s^2)\cos{2\phi}+{\cal K}\sin{2\phi}\right]\nonumber\\
&&\ \ \ \ \ \ \ \ \ \ \ \ \ \ \ \ \ -4sr_s\cos{2\beta}+t_s^2(1-s^2),\nonumber\\
D_1&=&-2t_s\alpha\sin{\phi}\,(1+sr_se^{2i\beta}),\nonumber\\
D_2&=&2t_s\alpha\cos{\phi}\,(1-sr_se^{2i\beta}).\nonumber
\end{eqnarray}
Here we set $\xi=0$ for simplicity. As is explained in Ref.~\cite{BC2}, the characteristics of the optical spring is defined in $M$. The equation $M=0$ is equivalent to
\begin{eqnarray}
&&\Omega^2(\Omega-\Omega_+)(\Omega-\Omega_-)\nonumber\\
&&\ \ \ \ \ \ +\frac{8\omega_0I_0}{mL^2}\frac{r_s\sin{2\phi}}{s(1+r_s^2)+(1+s^2)r_s\cos{2\phi}}=0
\end{eqnarray}
with
\begin{eqnarray}
\Omega_\pm=\frac{\gamma\left[-ist_s^2\pm\sqrt{4s^2r_s^2-(1+s^2)^2r_s^2\cos^2{2\phi}}\right]}{s(1+r_s^2)+(1+s^2)r_s\cos{2\phi}}.\\
\nonumber
\end{eqnarray}
In the absence of the opto-mechanical coupling ($I_0\rightarrow0$), the roots of the equation $M=0$ read $\Omega=0$ and $\Omega=\Omega_\pm$. The former represents the optical spring and the latter represents the optical resonances of the coupled cavity with the arm cavities and the signal-recycling cavity. When $2s/(1+s^2)$ becomes smaller than $\cos{2\phi}$, $\Omega_+$ makes pure imaginary and the dip for the optical resonance disappears. The imaginary part of $\Omega_+$ should be negative to keep the system stable:
\begin{eqnarray}
s(1+r_s^2)>\left|(1+s^2)r_s\cos{2\phi}\right|.\label{eq:stability}
\end{eqnarray}
The resonances shift with the opto-mechanical coupling, for which we can apply a perturbative analysis as is performed in Ref.~\cite{BC2}. The shift of the optical resonance is very small. By contrast, the shift of the optical spring resonance, $\Delta\Omega_0$ is significant:
\begin{eqnarray}
\Delta\Omega_0^2=-\frac{1}{\Omega_+\Omega_-}\frac{8\omega_0I_0}{mL^2}\frac{r_s\sin{2\phi}}{s(1+r_s^2)+(1+s^2)r_s\cos{2\phi}}.\label{eq:omega0}
\end{eqnarray}
Since $\cos{2\phi}$ can be negative, the denominator of the right hand side of Eq.~(\ref{eq:omega0}) could be zero or negative, in which case, however, the stability condition (\ref{eq:stability}) breaks. As far as the condition is satisfied, the optical spring frequency can be increased by tuning the squeeze factor $s$.

\begin{figure}[t]
		\includegraphics[scale=0.5]{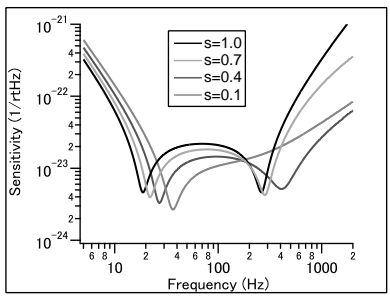}
		\caption{\label{fig:spectrum3}Quantum noise spectra of a detuned interferometer with the parametric amplifier. The amplifier changes the optical spring frequency with the squeezing factor $s$ as if the input power increases.}
\end{figure}

Figure~\ref{fig:spectrum3} shows the quantum noise spectra with the parametric amplifier in the signal recycling cavity. Here we extract the signal in the $b_2$ quadrature and the spectrum is given by taking a square root of $S_h=(|A_{21}^2|^2+|A_{22}|^2)/D_2$. With the squeezing factor increased, the optical spring frequency moves up and the optical resonance disappears at some point.

\section{Application of parametric amplifier}
Let us introduce two possible applications of the parametric amplifier for the gravitational-wave detector. One is to use it with the local readout scheme introduced in Ref.~\cite{LR}. While the gravitational wave cannot be measured from the local motion of the test mass in an interferometer without the optical spring, the test mass is physically driven by the gravitational wave via the optical spring in a detuned interferometer. The local motion measurement can be more sensitive as there is almost no constraint of the signal duration time in a short interferometer. The local readout can be more powerful with the amplifier that increases the optical spring frequency.

Another application is to use the amplifier with frequency dependent squeezing. Since the optical spring frequency is now a function of the squeezing factor and the squeezing angle and these parameters can be varied at different frequencies, the optical spring can be created in a broad frequency band. This idea, first proposed by Chen~\cite{ChenPhD}, is a realization of a so-called {\it optical lever} regime invented by Khalili~\cite{QM}.


\section*{Acknowledgement}
The authors would like to appreciate Prof. Yanbei Chen and Prof. Yuri Levin for valuable discussions. 
This research is supported by Japan Society for the Promotion of Science (JSPS).

\bibliographystyle{junsrt}
\pagestyle{headings}

\end{document}